\documentclass[12pt]{article}

\usepackage{epsf}

\usepackage{graphicx}%

\def\refitem#1{\relax}

 \newfont{\cyrfnt}{wncyr9 scaled 1120}

\begin{document}

\begin{center}
{\bfseries Recent Results of the  Hadron Resonance Gas Model  and the Chemical Freeze-out of Strange Hadrons}

\vskip 5mm

K. A. Bugaev$^{1 *}$,  A. I. Ivanytskyi$^{1}$, D. R. Oliinychenko$^{1, 2}$, E. G. Nikonov$^{3}$, V. V. Sagun$^{1}$
and G. M. Zinovjev$^1$

\vskip 5mm

{\small
$^1${\it
Bogolyubov Institute for Theoretical Physics, Metrologichna str. 14$^B$, Kiev 03680, Ukraine
}
\\
$^2${\it
FIAS, Goethe-University, Ruth-Moufang Str. 1, 60438 Frankfurt upon Main, Germany
}
\\
$^3${\it
Laboratory for Information Technologies, JINR, Joliot-Curie str. 6, 141980 Dubna, Russia
}
\\
$*$ {\it
E-mail: Bugaev@th.physik.uni-frankfurt.de
}}
\end{center}

\vskip5mm

\centerline{\bf Abstract}
{\small
A detailed discussion of  recent  results obtained 
within the hadron resonance gas model with the  multi-component  hard core repulsion is presented.
Among them there are the adiabatic chemical freeze-out criterion,  the concept of separate chemical freeze-out of strange particles and  the effects of   enhancement and  sharpening of wide resonances and quark gluon bags occurring in a thermal medium.
These findings are discussed in  order to strengthen   the planned heavy-ion collision  experimental programs  at 
low collision energies. We argue, that due to found effects,  at the center of mass collision energy 4-8 GeV the quark gluon bags may appear directly or in decays
as new heavy   resonances with the narrow width of about 50-150 MeV and with  the  mass above 2.5 GeV.
}

\vskip5mm

\begin{figure}[htbp]
  \centerline{
    \includegraphics[height=70mm]{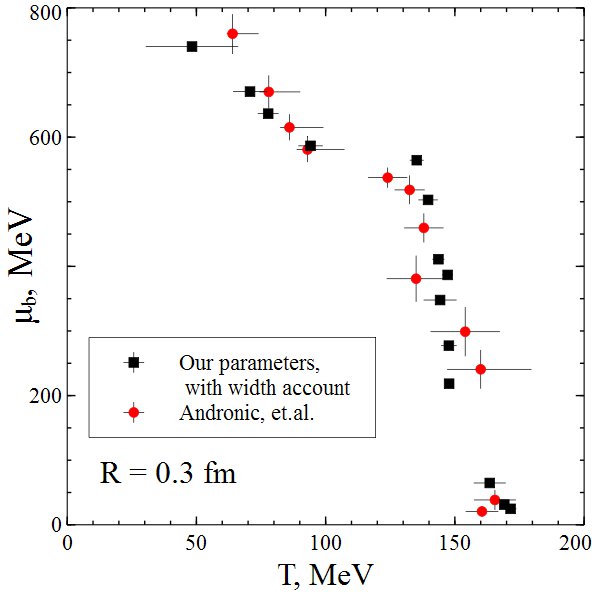}
  \hspace*{1.cm}
   \includegraphics[height=70mm]{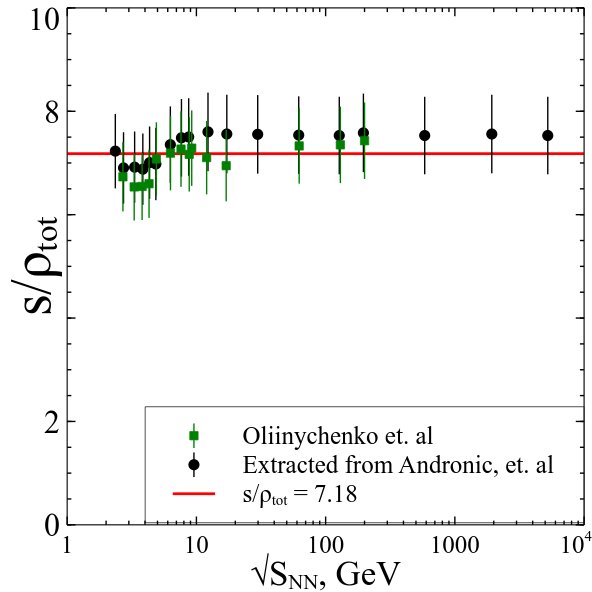}
  }
 \caption{Comparison of the chemical FO parameters of  HRGM1 \cite{KABAndronic:05} and  HRGM2 \cite{KABOliinychenko:12}. {\bf Left panel:} 
 baryonic chemical potential vs.  the 
 chemical FO temperature for HRGM1 (circles) and HRGM2 (squares).
{\bf Right  panel:} entropy per particle at chemical freeze-out $s/\rho \simeq 7.18$ 
vs.  the center of mass energy per nucleon $\sqrt{s_{NN}}$.
 The shown errors  are combined the statistical and systematic errors.
The results of  HRGM2 (squares) are  very similar to  that ones obtained by HRGM1.}
  \label{Fig1}
\end{figure}

{\bf 1. Basic elements of  the hadron resonance gas model.} The recent findings \cite{KABOliinychenko:12,HRGM:13,ResModif:12} obtained by  the hadron resonance gas model (HRGM)  resolved a few old puzzles on the chemical freeze-out and gave a novel look at some old problems of QCD phenomenology. 
One of the most  successful version of  the HRGM, the HRGM1,  was developed in \cite{KABAndronic:05,KABAndronic:09}.
The novel version of this model, the HRGM2, which was worked out in \cite{KABOliinychenko:12,HRGM:13}, has 
the same basic properties as the HRGM1, namely, it employes the hard-core repulsion between the hadrons,  
it includes all hadronic resonances with masses up to 2.5 GeV with their finite width. Also it accounts both the thermal  hadronic multiplicities and the nonthermal  ones which are coming from the decay of heavy resonances. Finally, 
the  HRGM2 explicitly  accounts  for  the strangeness conservation  by  finding out the chemical potential of the strange charge from a condition of vanishing strangeness.  Despite slightly different particle tables employed  in the HRGM1 and HRGM2  and a systematic accounting  for the  resonance width in HRGM2, 
these models  give very similar results  for the chemical freeze out  (FO) parameters  for the same values of the hadronic hard-core radii  
\cite{KABOliinychenko:12}. This  is shown  in Fig. \ref{Fig1} for the same hadronic hard-core radii  $R = 0.3$ fm.   

\begin{figure}[htbp]
  \centerline{
    \includegraphics[height=70mm]{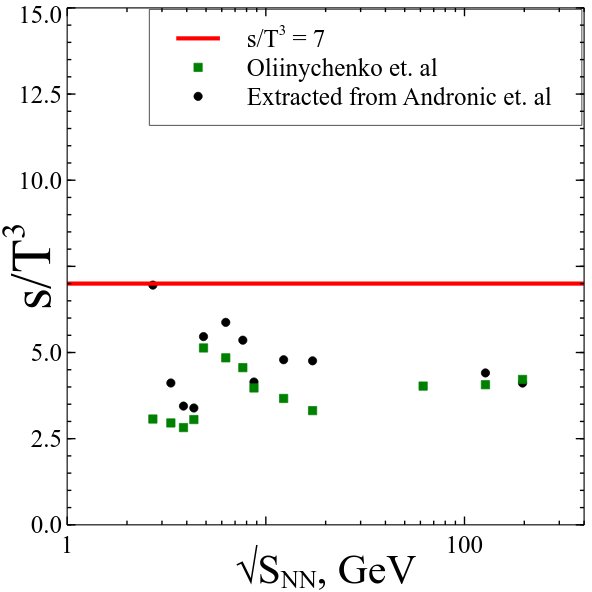}
  \hspace*{1.cm}
   \includegraphics[height=70mm]{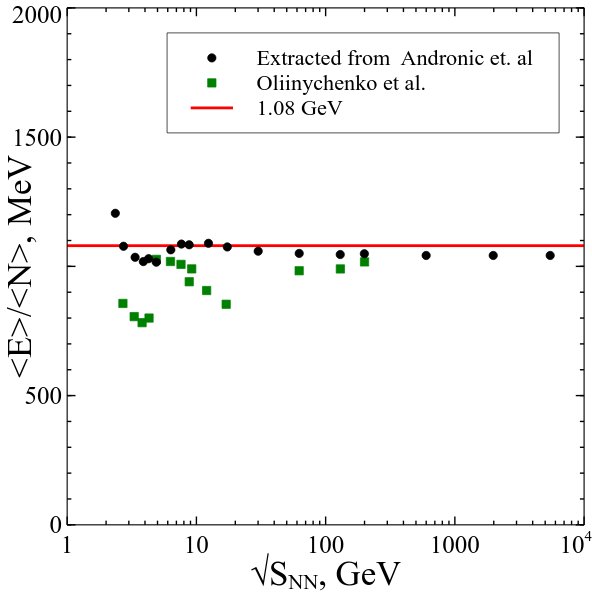}
  }
 \caption{Different chemical freeze-out criteria. {\bf  Left panel:} ratio of the entropy density to the cube of temperature $s/T^3$ at chemical freeze-out vs. the center of mass energy $\sqrt{s_{NN}}$. Results of  the HRGM1  (HRGM2)
 are shown by circles (squares). 
{\bf  Right  panel}: energy per particle  $\langle E \rangle / \langle N \rangle$  at chemical chemical freeze-out vs. $\sqrt{s_{NN}}$. Notations are the same as in the left panel.
}
  \label{Fig2}
\end{figure}

{\bf 2. Adiabatic chemical FO criterion and effective hadronic mass spectrum.}
A thorough investigation of the traditional chemical FO criteria performed in  \cite{KABOliinychenko:12} for the  HRGM1 and   HRGM2 gave rather valuable results.  Although   the discussion  about the reliable chemical FO criterion has a long history \cite{KABAndronic:05,Cleymans:06}, only  very recently   it was demonstrated that none of the previously suggested chemical FO criteria, including the most popular one of constant energy per particle $E/N \simeq 1.1$ GeV  \cite{KABCleymansFO,KABCleymansFO:b}
and the criterion of constant  entropy density $s$ to the cube of  FO temperature ratio, $s/T^3 \simeq 7 $ \cite{Tawfik:06a,Tawfik:06b},  is  robust \cite{KABOliinychenko:12},  if the realistic particle table with the hadron masses up to 2.5 GeV is used and the hard-core hadronic repulsion is included.  This is clearly seen by comparing  the 
right panel of Fig. \ref{Fig1} with  both panels of  Fig. \ref{Fig2}. 
At the same time in \cite{KABOliinychenko:12} 
 it was  shown that despite an essential difference with the approach used in \cite{KABAndronic:05},
 the both versions of the hadron resonance gas model demonstrate almost the same value 7.18  for   the  entropy per particle  at chemical FO. In other words, the criterion of adiabatic chemical FO  is, indeed, the robust  one. 

A puzzle of  the adiabatic  chemical  FO led us to a formulation of the model equation of state \cite{ResModif:12,ResModif:13}
\begin{eqnarray}\label{EqI}
&&p_M= C_M ~ T^{A_M} ~ \exp\left[- \frac{m_M}{T}\right] \,, \quad 
p_B= C_B ~\, T^{A_B}\, ~ \exp\left[ \frac{\mu_B- M_B}{T}\right] \,,
\end{eqnarray}
which successfully  parameterizes the baryonic pressure $p_B$  and the mesonic one $p_M$  of the HRGM2.  
The pressure of antibaryons is also described by the right Eq. (\ref{EqI}), but for negative value of the baryonic chemical potential  extracted from the data at chemical FO, i.e. $\mu_{\bar B} = - \mu_B$.  
The constants
$C_a$, $A_a$, $m_a$  with $a \in \{M, B \}$  in (\ref{EqI})  parameterize the integrated hadronic mass spectrum  and  should 
be determined  from the fit of  the adiabatic chemical FO parameters. 
The remarkable fact, however, is that one can get rather good description of the constant entropy per particle by fitting  only the
 mesonic and baryonic particle densities for which the mean deviations  squared per degree of freedom, respectively, are $\chi_M^2/dof \simeq 0.42/11$ and  $\chi_B^2/dof \simeq 7.8/11$. In other words, once the particle densities of  the system (\ref{EqI}) are reproduced,
 the corresponding entropy densities are automatically  reproduced well giving the total  mean deviation squared $\chi_{tot}^2/dof \simeq 19.4/36$. The latter includes the sum of  mean deviations squared of meson density together with the one of baryons and their full entropy per particle  at 
 chemical FO. 

The system (\ref{EqI}) is a further refinement of  the  original E. Shuryak idea proposed for vanishing baryonic densities \cite{EffEOS:73} and the found  powers $A_B \simeq 6.097 \pm 0.38$ and  $A_M \simeq  5.31 \pm 0.14$ are very close to the early estimate of Ref.  \cite{EffEOS:73}. Additionally,  under the well justified assumptions  the present model allows us to uniquely determine an effective density of states of hadronic mass spectrum   using the inverse Laplace technic \cite{ResModif:12}. The found density of  baryonic states is   $\frac{\partial   \varrho_B (m) }{\partial m} \sim \frac{(m-M_B)^{A_B-3.5}}{  m^{\frac{3}{2}} }$, while the mesonic one is
$\frac{\partial   \varrho_M (m) }{\partial m} \sim  m^{A_M-4}$, i.e. these densities of states are power-like rather than the exponential one.
Moreover,   numerical estimates show that 
 for hadronic masses below $2.5$ GeV these densities of states   essentially differ from   the empirical  power-like  hadronic  density of states $\frac{\partial \varrho^{emp} (m) }{\partial m} \simeq  \frac{4.11}{470\, \rm MeV}\,  \left[ \frac{m}{470\, \rm MeV}\right]^{3.11}$ found in \cite{KABHagedorn4} from the  Particle Data Group  \cite{KABPDG:08} data.  Our  searches for   possible  reasons of such a  difference led  to 
a  discovery of two new effects based on the modification of  the wide resonance (and quark gluon bag)  properties  in a thermal medium \cite{ResModif:12,ResModif:13}.  

\begin{figure}[htbp]
  \centerline{
    \includegraphics[width=77mm]{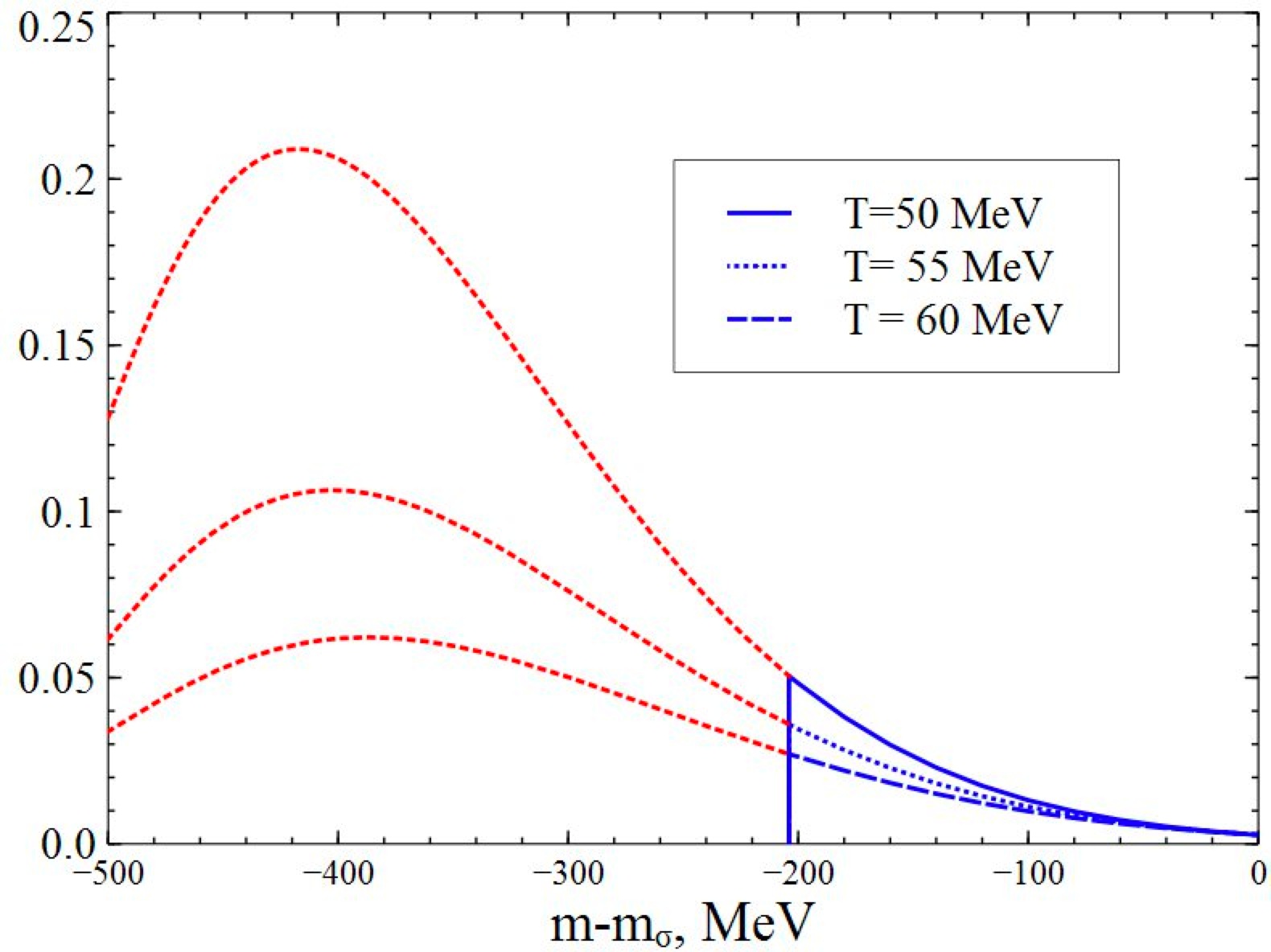}
  \hspace*{-0.22cm}
   \includegraphics[width=77mm]{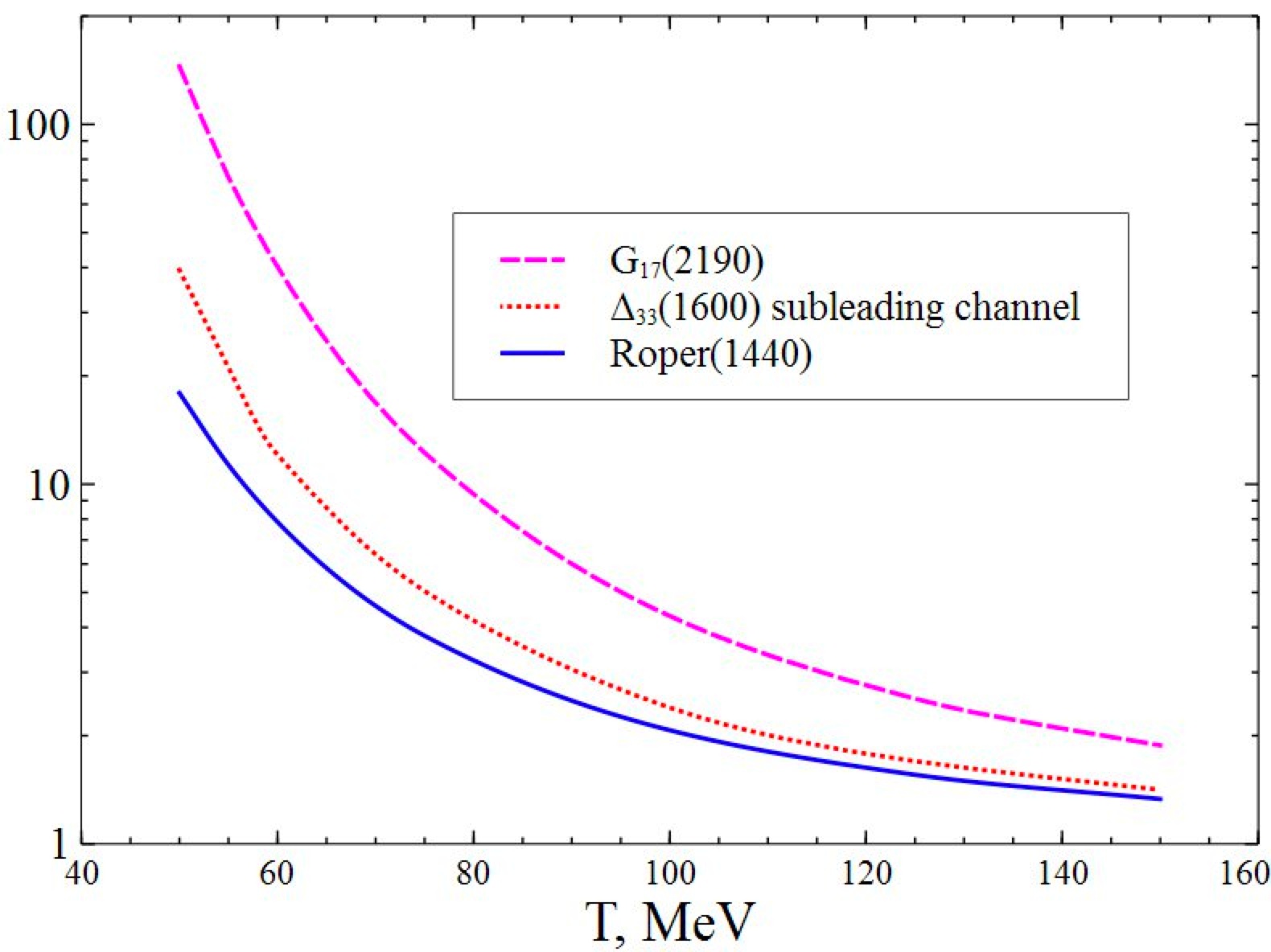}
  }
 \caption{{\bf Left panel:}
 Typical temperature dependence of  the mass distribution  $f_\sigma (m)/ \phi(m_\sigma, T)  $ (in units of $1/$MeV, see Eq. (\ref{EqIII})) for $\sigma$-meson with the mass $m_\sigma = 484$ MeV, the width $\Gamma_\sigma = 510$ MeV  \cite{KABSigma:07} and two pion threshold $M_\sigma^{Th} = 2 \, m_\pi \simeq 280$ MeV.  
 In the left panel the short dashed curves demonstrate that at these temperatures  the maximum of  the full mass distribution is shifted well below the two pion threshold (vertical line at $m-m_\sigma = -204$ MeV)
 and, hence, the resulting effective mass attenuation of a wide resonance acquires an icy slide shape.
{\bf Right panel:} Typical temperature dependence of  the resonance  enhancement. The ratio $R(T) = \frac{F_k}{g_k\, \phi (m_k,T)}$  is shown for a few hadronic resonance decays.  For wide resonances the effect of enhancement can be huge. 
}
  \label{Fig3}
\end{figure}

{\bf 3.  Thermal  enhancement and  sharpening of wide resonances at chemical FO.}
In order to demonstrate these new effects, here we  consider the Gaussian mass attenuation  of wide resonances instead of the Breit-Wigner one that is used in the  actual  simulations,  since in this case  the evaluation is more transparent.   
Such a treatment gives only about 10 \% difference from  the Breit-Wigner one \cite{ResModif:13}, but it also  allows us to obtain some important conclusions on the mass spectrum of 
quark-gluon  (QG) bags which according to \cite{FWM:08,FWM:09} should unavoidably have the Gaussian mass attenuation. Note    that  such an estimate  provide us with the lower limit,
since  the Gaussian  mass distribution  vanishes  much faster than the Breit-Wigner one. The typical term of  the $k$-resonance  that enters into  the  mass spectrum of  the hadron resonance gas model   is given by  $ F_k (\sigma_k) \exp{ \left[ \frac{\mu_k}{T} \right] }$ \cite{KABOliinychenko:12,KABAndronic:05} (here $\mu_k$ is the chemical potential of this resonance) with
\begin{eqnarray}
\label{EqII}
%
F_k (\sigma_k) & \equiv &  g_k \int\limits_{0}^\infty  d m \,  \frac{\Theta\left(  m - M_k^{Th} \right) }{N_k (M_k^{Th})} 
 \, \exp \left[ - \frac{(m_k -m)^2}{2\, \sigma_k^2}  \right] 
 \int \frac{d^3 p}{ (2 \pi \hbar)^3 }   \exp \left[ -\frac{ \sqrt{p^2 + m^2} }{T} \right]  \,. \quad \quad 
\end{eqnarray}
Here $m_k$ is the mean mass of the $k$-th resonance, $g_k$ is its degeneracy factor, $\sigma_k$ is the Gaussian width which is related to the true resonance width  as $\Gamma_k = Q\,  \sigma_k$ (with $Q \equiv 2 \sqrt{2\, \ln2}$) and the normalization factor is defined via the threshold mass $M_k^{Th}$   of the dominant channel as $N_k (M_k^{Th}) \equiv  \int\limits_{M_k^{Th}}^\infty  d m \, 
 \, \exp \left[ - \frac{(m_k -m)^2}{2\, \sigma_k^2}  \right] $.
For the narrow resonances the term $F_k (\sigma_k)$ converts into the usual thermal density of particles, i.e. for $\sigma_k \rightarrow 0$ one has $F_k \rightarrow  g_k \, \phi(m_k, T)$, where the following notation is used $ \phi(m,  T) \equiv  \int \frac{d^3 p}{ (2 \pi \hbar)^3 } \exp \left[ -\frac{ \sqrt{p^2 + m^2} }{T} \right] $.  
The momentum integral in (\ref{EqII}) can be written using the non-relativistic approximation $\phi(m, T) \simeq
\left[ \frac{m\, T}{2\, \pi \, \hbar^2} \right]^\frac{3}{2}\exp \left[   -\frac{  m } {T} \right] $. Then  to simplify the mass integration of  (\ref{EqII})  one can make the full square in it from the  powers of  $(m_k -m)$ and get  
\begin{eqnarray}
\label{EqIII}
\hspace*{-0.22cm}
F_k (\sigma_k) & \equiv & g_k  \int\limits_{0}^\infty  d m \, f_k ( m)  \simeq   \tilde g_k    \int\limits_{0}^\infty  d m \,  \frac{\Theta\left(  m - M_k^{Th} \right)}{N_k (M_k^{Th})}
 \, \exp \left[ - \frac{(\tilde m_k -m)^2}{2\, \sigma_k^2}  \right]  
 \left[ \frac{m\, T}{2 \pi   \hbar^2} \right]^\frac{3}{2}\exp \left[   -\frac{  m_k } {T} \right]  \,, ~
\end{eqnarray}
where  an effective resonance  degeneracy $\tilde g_k$   and   an effective resonance mass $\tilde m_k$ are defined as
\begin{eqnarray}\label{EqIV}
\tilde g_k     & \equiv &  g_k  \exp \left[  \frac{\sigma_k^2}{2 \, T^2} \right] = g_k  \exp \left[  \frac{\Gamma_k^2}{2 \,Q^2\, T^2} \right] \, , \quad 
 \tilde m_k   \equiv  m_k - \frac{ \sigma_k^2 }{T} = m_k - \frac{ \Gamma_k^2 }{Q^2\, T} \,.
\end{eqnarray}
From Eq.  (\ref{EqIV})
one can  see that  the presence of the width, firstly,  may  strongly  modify the degeneracy  factor $g_k$ and,
 secondly,  it may essentially shift the maximum of the mass attenuation towards  the threshold or even below it.
 There are two corresponding effects which we named as {\it the near threshold thermal resonance enhancement} and  {\it the near threshold resonance sharpening}.   
 These effects formally appear due to the same reason as the  famous Gamow window for the thermonuclear reactions in stars \cite{KABGamow1}:
 just above the resonance decay threshold 
 the integrand $f_k(m)$ in (\ref{EqIII}) is a product of two functions of a virtual resonance mass $m$, namely, the Gaussian attenuation is an increasing 
 function of $m$, while the Boltzmann exponent strongly decreases above the threshold. The resulting attenuation of their product has a maximum,
 whose shape, in contrast to the usual Gamow window,  may be extremely asymmetric due to a threshold presence.  Indeed, 
 as one can see from the left panel of  Fig. \ref{Fig3} the resulting mass attenuation of a  resonance may acquire  the form of  the sharp and narrow peak that  
 closely resembles 
 an icy slide. 
 
To our best knowledge a special attention to 
the thermal enhancement  of the $\Delta_{33} (1232)$ isobar   in a thermal medium was  for the first time paid  in \cite{KABStashek:87}, although no detailed explanation of this effect was given in \cite{KABStashek:87}. 
 The  mass shift  and shape modification  of  $\rho$ meson  in a hot hadronic matter   were  found   in  
 \cite{KABBarz:1991}. However, in both cases  the found effects were not very strong since the considered resonances are not really wide. An analysis of  the  wide resonances' modification   performed in \cite{ResModif:12,ResModif:13} gave  the astonishing   results.  
 From   the effective resonance degeneracy $\tilde g_k$  and 
 the effective resonance mass $\tilde m_k$  defined  in  (\ref{EqIV})  one can see that  they  essentially differ from their vacuum values  for $T \ll \sigma_k$. 
 The left panel of  Fig. \ref{Fig3}  demonstrates a strong modification of  the $\sigma$-meson mass attenuation at low temperatures which leads to the near threshold resonance sharpening. 
 A simple analysis  shows that the effect of resonance sharpening is strongest, if the threshold mass is shifted above  the convex part of the Gaussian distribution in (\ref{EqIII}), i.e. for $M_k^{Th} \ge \tilde m_k$ or for the temperatures  $T$ well  below $T^+_k \equiv \frac{\sigma_k^2}{m_k - M_k^{Th}}$.  In this case near the threshold  a resonance acquires  a narrow effective  width   \cite{ResModif:12,ResModif:13}
\begin{eqnarray}
\label{EqV}
\Gamma_k^N (T)  & \simeq  &     \frac{\ln(2)}{ \frac{1}{T} -  \frac{1}{T^+_k} }    \,.  
\end{eqnarray}
The right panel of  Fig. \ref{Fig3} shows  that  an effect of  near threshold enhancement compared to the thermal particle density of the same resonance taken with a vanishing width 
  can be, indeed,  huge for wide ($\Gamma \ge  450$ MeV) and medium wide ($\Gamma \simeq 300-400$ MeV) resonances. This effect can naturally explain the strong temperature dependence of  hadronic pressure 
(\ref{EqI})  at chemical FO, which in its turn generates the discussed  power-like  mass spectrum of  hadrons.
 The other important conclusion from this analysis  is that there is no sense to discuss the mass spectrum of 
hadronic resonances, empirical or Hagedorn, without a treatment of  their width. Furthermore, the same is true 
for  the QG  bags which, according to the finite width model \cite{FWM:08,FWM:09},  are heavy and wide  resonances 
having the  mass $M_B$ larger than $M_0 \simeq 2.5$ GeV and  the mean width of the form $\Gamma_B \simeq \Gamma_0 (T) \left[  \frac{M_B}{M_0}  \right]^\frac{1}{2}$.  Here  $\Gamma_0 (T)$ is a monotonically increasing function  of temperature  $T$ and  $\Gamma_0 (0) \in [400; 600]$ MeV
\cite{FWM:08,FWM:09}.  

The estimates of  \cite{ResModif:12,ResModif:13} based on the finite width model and the approach outlined above
show that our  best  hopes  to find  the QG bags  experimentally  may be related to their sharpening and enhancement  by  a thermal medium. Then at the  chemical FO temperatures of about $T  \simeq  80-140$ MeV  the {\it QG bags may appear directly or in decays
as narrow  resonances with  the width of about 50-150 MeV which have  the  mass about or above 2.5 GeV and  which  are absent in the tables of elementary particles.}  
 Note that this range of  chemical FO temperatures  correspond to the center of mass  energy of collision $\sqrt{s_{NN}} \in [4; \, 8]$ GeV \cite{KABOliinychenko:12,KABAndronic:05}, which is in the range of the  NICA   JING and FAIR GSI energies of collision.
This energy range sets the  most promising  kinematic limit for the QG bag searches. 
\begin{figure}[htbp]
\centerline{
          \includegraphics[height=7 cm]{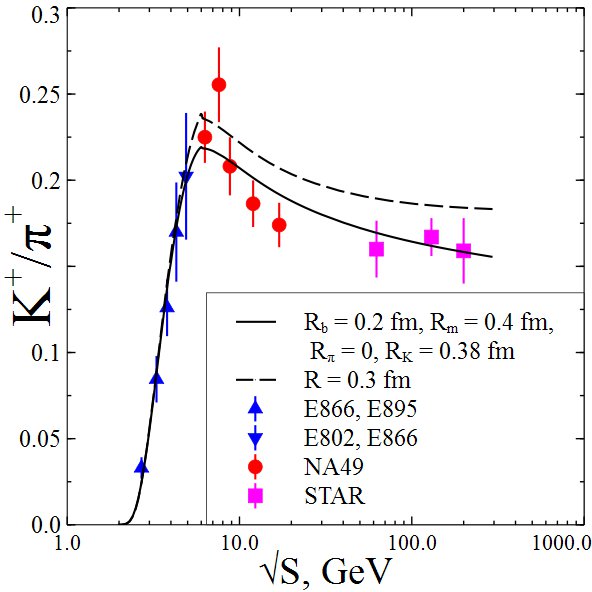}   
	\hspace*{1.cm}
	\includegraphics[height=7.cm]{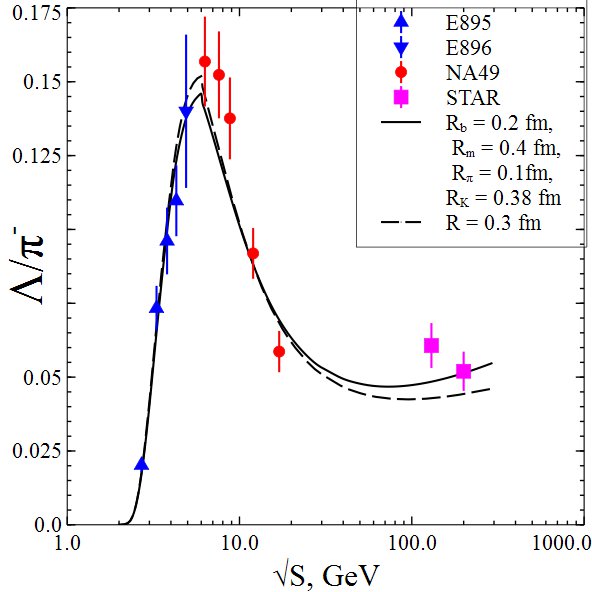} 
}
 \caption{$\sqrt{s_{NN}}$  dependences  of  $K^+/\pi^+$  (left panel)
 and  $\Lambda/\pi^-$ (right panel) ratios 
  obtained  in  \cite{HRGM:13}  within  the multicomponent model are compared to that ones 
  found within the  one-component model  \cite{KABOliinychenko:12}. 
  }
  \label{Fig4}
\end{figure}  
 
{\bf 4.  Strangeness Horn description and chemical FO of strange particles.} A  novel feature of principal importance  implemented into the  HRGM2  \cite{KABOliinychenko:12,HRGM:13} is  its multicomponent hard-core repulsion.  
One of the traditional difficulties  of the HRGM was related to the  Strangeness Horn description  which up to recently was far from being satisfactory, although very different versions of  the HRGM were used for this purpose (see \cite{HRGM:13} for a discussion and references).  The HRGM2 with the multicomponent 
hard-core repulsion  allows one   to describe the hadron yield ratios from the low AGS to the highest RHIC energies. In \cite{HRGM:13} it was  demonstrated  that the variation of the hard-core radii of pions and kaons leads  to a drastic improvement of   the fit quality of the measured  mid-rapidity data  and for the first time 
such a model provides us with 
 a complete    description of   the Strangeness Horn behavior as the function of the energy of collision without spoiling the fit quality of other ratios. The  best global fit with $\chi^2/dof  = 80.5/69 \simeq 1.16$ is found for  an almost vanishing hard-core radius of pions of 0.1 fm and for the hard-core radius of  kaons being equal to 0.38 fm, whereas the hard-core radius of all other mesons is found to be  0.4 fm and that one of baryons is equal to  0.2 fm.  Fig. \ref{Fig4} demonstrates that within the multicomponent HRGM2  the mean  deviation squared per data  is  $\chi^2/dof \simeq 7.5/14$ for   $K^+/\pi^+$  and  $\chi^2/dof \simeq 14/12$ for  $\Lambda/\pi^-$  ratios. Note that the   value  of $\chi^2/dof $ obtained  for  $K^+/\pi^+$ ratio   is almost three times smaller  compared to the best fit of the  one-component model.
\begin{figure}[htbp]
\centerline{
          \includegraphics[height=7 cm]{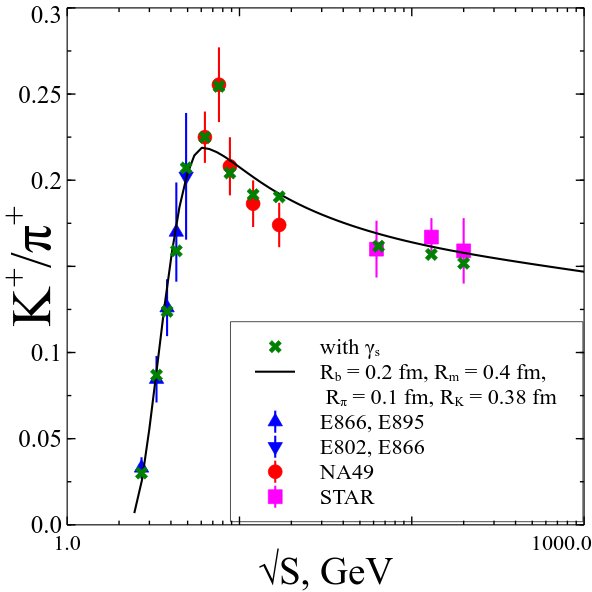}   
	\hspace*{1.cm}
	\includegraphics[height=7.cm]{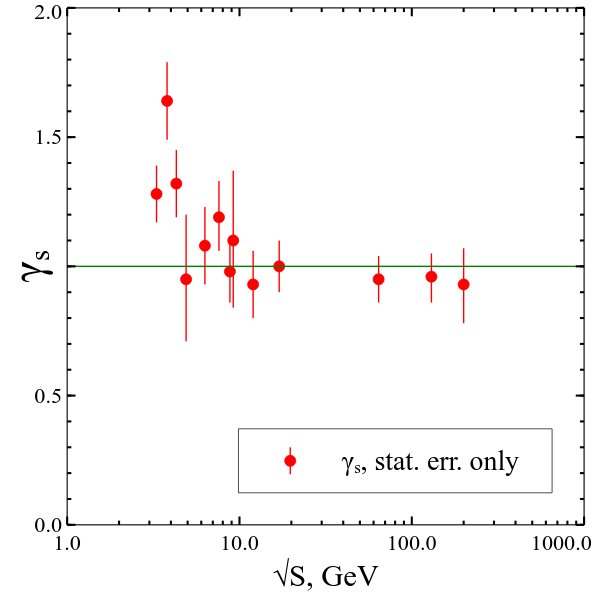}
	}
\centerline{
          \includegraphics[height=7 cm]{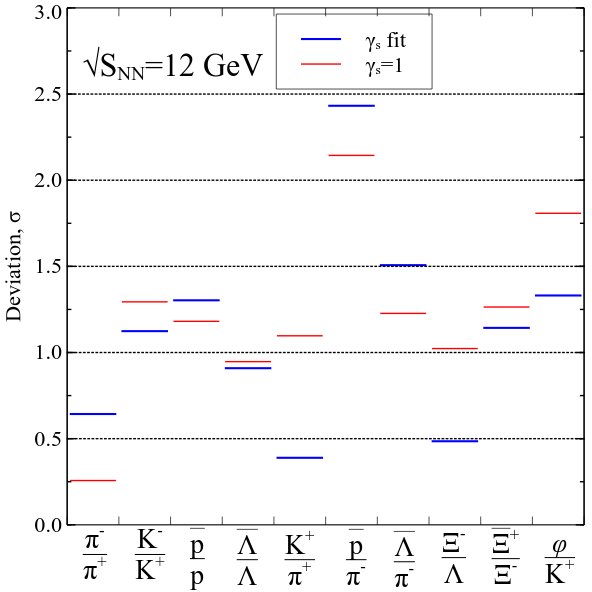}   
	\hspace*{1.cm}
	\includegraphics[height=7.cm]{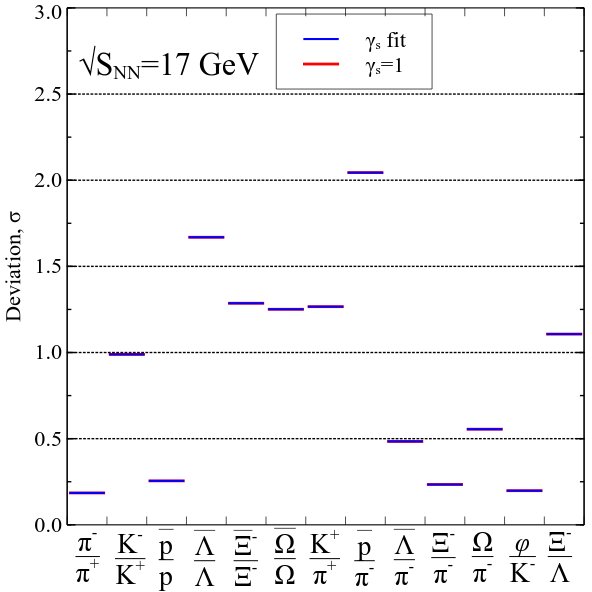}}	

 \caption{{\bf Upper left panel:}$\sqrt{s_{NN}}$  dependence   of  $K^+/\pi^+$ ratio  for  $\gamma_s$ included into the fit (crosses) and for  $\gamma_s = 1$ (solid curve) \cite{HRGM:13}.
 {\bf Upper right panel:} $\sqrt{s_{NN}}$  dependence  of  $\gamma_s$  obtained from the new fit shown in the left panel.  {\bf Lower  panels:} Comparison of the relative deviations of the fit results from experimental ratios 
is shown for  $\sqrt{s_{NN}} \simeq 12$ GeV (left) and  $\sqrt{s_{NN}} \simeq 17$ GeV (right).
  }
  \label{Fig5}
\end{figure}

Note, however, that  in order to demonstrate its new abilities  the strangeness suppression factor  $\gamma_s$ \cite{StrangSuppr:03,StrangSuppr:05} in the  multicomponent HRGM2 \cite{HRGM:13} simulations  was kept to be a unity.   In the most recent version of the HRGM2 \cite{HRGM2:13} we fixed  the hard-core radii values found earlier and 
included $\gamma_s$ into the fitting procedure.  The obtained results have a  better  fit quality  $\chi^2/dof  = 63.4/55 \simeq 1.15$ for all 14  energies of collision in the range from the low  AGS to the highest RHIC energies.  
As one can see from the upper left panel of  Fig. \ref{Fig5}  the $\gamma_s$ fit greatly improves the 
$K^+/\pi^+$ ratio allowing us to correctly reproduce even  the peak of the Strangeness Horn! From the lower panels of this figure one, however, can deduce that the $\gamma_s$ fit does not allow us to improve the description of  the ratios
of   antihyperon-hyperon with nonzero strangeness, i.e.   $\bar \Lambda/\Lambda$, $\bar \Xi/\Xi$ and $\bar \Omega/\Omega$.  Also  the antiproton to pion ratio is  almost not improved by this fit.  Moreover, the  $\gamma_s$ fit does not improve any ratios at $\sqrt{s_{NN}} \simeq 17$ GeV (see the lower right panel in Fig. \ref{Fig5})!

In order to get rid of the phenomenological parameter  $\gamma_s$ and to further improve the fit quality of  hadron multiplicities, we worked out an alternative concept  of  the strange particles chemical  FO (SPFO) \cite{HRGM2:13}. Its main assumption is that the SPFO occurs separately from the  chemical FO of nonstrange hadrons (UDFO hereafter). A partial  justification for such a hypothesis  is given in  \cite{EarlyFO:1,EarlyFO:2,EarlyFO:3}, where the early chemical and kinetic FO of  $\Omega$ hyperons  
\begin{figure}[htbp]
\centerline{
          \includegraphics[height=7 cm]{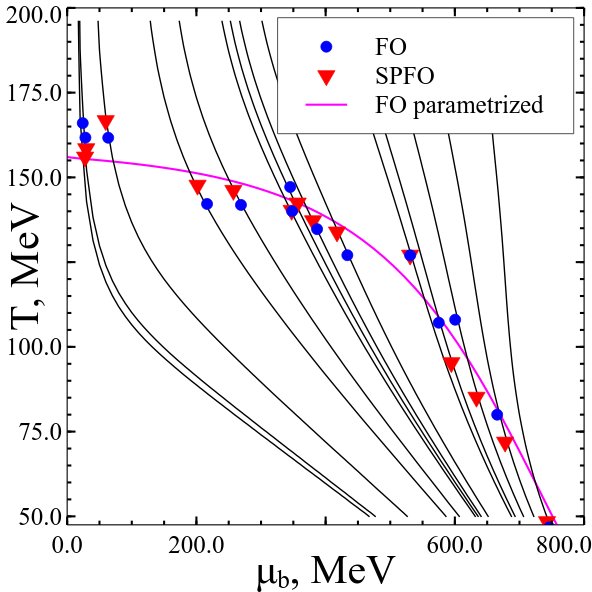}
	\hspace*{1.cm}
	\includegraphics[height=7.cm]{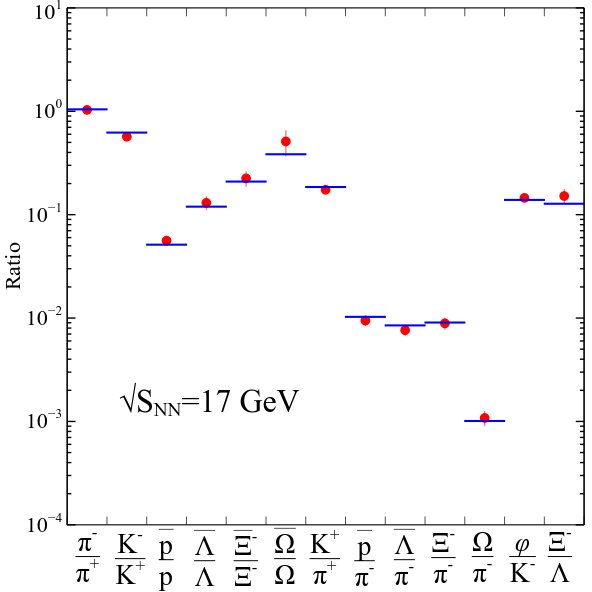}   
	}
\centerline{
	\includegraphics[height=7 cm]{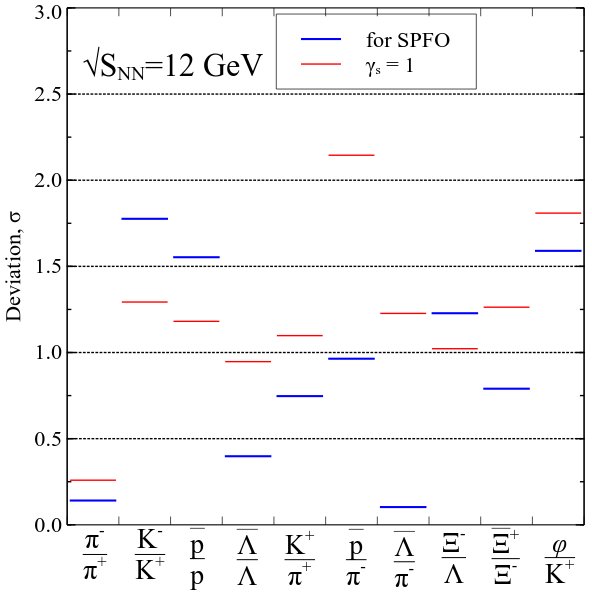}   
	\hspace*{1.cm}
	\includegraphics[height=7.cm]{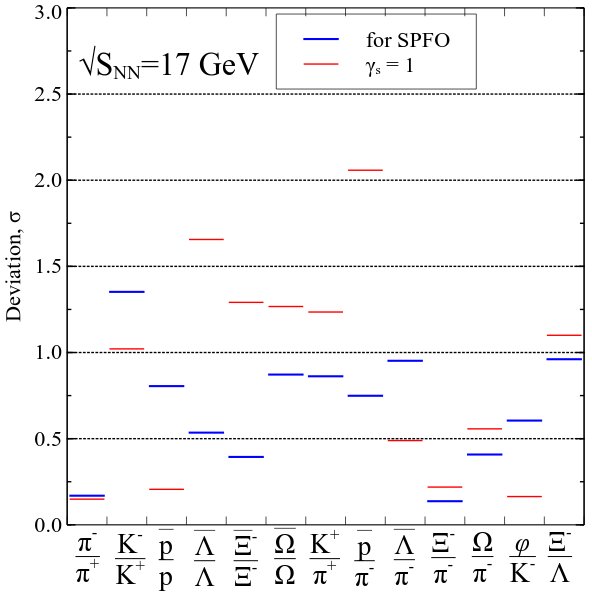}	
	}	

 \caption{{\bf Upper left panel:} $\mu_b$ and $T$ values for   UDFO (circles) and SPFO (triangles) are shown along with their isentropic trajectories  (thin curves).
 {\bf Upper right panel:} Example of the ratios description  within the  SPFO approach for $\sqrt{s_{NN}} \simeq 17$ GeV. {\bf Lower  panels:} Same as in the lower panels of  Fig. \ref{Fig5}, but for the SPFO approach.
  }
  \label{Fig6}
\end{figure}
 and  
 $J/\psi$ mesons is discussed for the energies above the highest SPS energy.

A similar idea of separate chemical FO of strange particles was suggested in \cite{Gupta13}.
 However, in contrast of ideal gas treatment of   \cite{Gupta13}, our  SPFO  concept  accounts for conservation laws
 connected the UDFO and the SPFO.  To determine the SPFO   baryonic chemical potential, the chemical potential for a  third projection  of isospin and strange chemical potential we employ an isentropic evolution 
of the system between  the UDFO and the SPFO, and the conservation laws of baryonic charge, third projection of isospin and strangeness. Therefore,  the SPFO  concept has  only one additional fitting parameter,  the SPFO temperature,
compared to the  HRGM2  \cite{HRGM:13} and, hence,  the number of its  independent degrees of freedom 
coincides with the one for the $\gamma_s$ fit. 

 The ratio of   hadron $p1$  multiplicity to  hadron $p2$  multiplicity is defined as
\begin{eqnarray}\label{EqVI}
\frac{n_{p1}^{final}}{n_{p2}^{final}} = 
\frac{\sum_{X} n_X^{therm}(SPFO) Br(X\to p1)\frac{V_{SPFO}}{V_{UDFO}} + \sum_{Y} n_Y^{therm}(UDFO) Br(Y\to p1)}
     {\sum_{X} n_X^{therm}(SPFO) Br(X\to p2)\frac{V_{SPFO}}{V_{UDFO}} + \sum_{Y} n_Y^{therm}(UDFO) Br(Y\to p2)} \,,
\end{eqnarray}
where the UDFO  and SPFO `volumes' (in fact, the corresponding  hypersurface extents) are related via the baryonic 
$n_B$ and isospin $n_{I3}$  densities  
as 
$\frac{V_{SPFO}}{V_{UDFO}} = \frac{n_B^{UDFO}}{n_B^{SPFO}} = \frac{n_{I3}^{UDFO}}{n_{I3}^{SPFO}}$ due to  conservation laws discussed above.  The sums in Eq. (\ref{EqVI}) run over over all hadronic 
species including $p1$ and  $p2$. In the latter case  the corresponding  branching ratios of decay are defined as  
$Br(p1\to p1) =1$  and  $Br(p2\to p2) =1$, whereas for  $X \neq  p$ the branching ratios of decay $Br(X\to p)$ are taken from the Particle Data Group \cite{KABPDG:08}. Due to the conservation laws  an  effective number of degrees of freedom 
in the SPFO approach is the same as in case of  the $\gamma_s$ fit, but the fit quality  is  better  $\chi^2/dof  =  58.5/55 \simeq 1.06$ than for the  $\gamma_s$ fit.
As one can see from Fig.  \ref{Fig6}, the fit of  particle ratios  which were problematic for the $\gamma_s$ fit
gets better, but not at an expense of  an essential worsening of other ratios. Therefore, {\it the SPFO concept \cite{HRGM2:13}, outlined here,
provides us with  more realistic and reliable approach, then a popular concept  of strange suppression factor $\gamma_s$.}

{\bf 5. Conclusions.} In this work we discussed the new results obtained within the multicomponent 
version of the HRGM2.  This model allowed us to demonstrate that the most robust criterion of chemical FO 
is the  adiabatic FO criterion of constant entropy per particle equal to $7.18$.  Searches for  a physical explanation 
of this phenomenon, led us to a thorough analysis of effective hadronic mass spectrum. One of the most important conclusions of our studies is that the width of resonances and QG bags   should unavoidably be included into statistical models and, hence, 
 there is no sense to discuss the hadronic mass spectrum, empirical or Hagedorn,
without accounting for the width of  constituents. Moreover, we showed that the properties of  wide resonances are essentially modified at chemical FO. Using the effect of near threshold resonance sharpening, we argued that, if the QG bags are formed in heavy ion collisions, then in experiments  they may  appear  directly or in decays as narrow and heavy  resonances with  the width about 50-150 MeV which are absent in the tables of elementary particles.  Our estimates show that 
such resonances  may appear at 
the chemical FO temperatures of about $T  \simeq  80-140$ MeV which corresponds to $\sqrt{s_{NN}} \in [4; \, 8]$ GeV. This  range of  the center of mass  energy of collision exactly fits into the NICA and FAIR Projects. 

Also we demonstrated that the HRGM2 with the multicomponent  hard-core repulsion is able to drastically improve the 
description of  $K^+/\pi^+$ and  $\Lambda/\pi^-$ ratios which were problematic for other formulations of HRGM.
The most recent  development  of the HRGM2 includes a novel concept  of the SPFO, which seems to be  physically  more adequate for a  description of  the strange  hadron multiplicity ratios than the popular strangeness suppression factor approach. Therefore, one of the future  physical tasks will be to find a physical  reason of  why  the chemical FO of strange particles  at   $\sqrt{s_{NN}} < 5$ GeV occurs at lower temperatures than the FO of  hadrons  built up  from
$u$ and $d$ quarks, and why  it is vise versa for $\sqrt{s_{NN}} >  9$ GeV. 

All versions of  the HRGM2 discussed here show that  in a narrow range of collision energy $\sqrt{s_{NN}} \in [4.3; 4.9]$ GeV there exist some peculiar  irregularities   in a behavior of the chemical FO temperature, in entropy density and in total particle  number density.  From the left panel of Fig. \ref{Fig1} it is seen that about 15 \% increase of $\sqrt{s_{NN}}$ leads to about 10 \% change  of the baryonic chemical potential $\mu_B$, whereas the chemical FO temperature increases by a  factor 1.5.  From the left panel of Fig. \ref{Fig2} one can also see a simultaneous  
appearance of 
irregularity for  $s/T^3$ ratio which  jumps over  2 times, leading to  6.75 times resulting increase  of   entropy density
while $\sqrt{s_{NN}}$ increases by about 15 \% only! Moreover, due to the adiabatic chemical  FO   such a  small variation of  collision energy  leads to about 
6.75 times increase of the  particle density, while  the  chemical FO `volume'  changes on  about 20 \%  \cite{KABOliinychenko:12} only!
Note that all these irregularities can be naturally  explained  within the shock adiabat model of central nuclear collisions \cite{SA1,SA2} as  a formation of the mixed quark-gluon-hadron phase.
However, 
to safely reveal the physical  cause  of all discussed phenomena   occurring at  the collision energy  interval $\sqrt{s_{NN}} \in [4.3; 4.9]$ GeV and to elucidate  their  relation to  the  irregularities observed   at slightly higher collision energy  $\sqrt{s_{NN}} \simeq 7.6$ GeV,   we  need an accelerator   of new generation working in this   energy range with much higher experimental  accuracy.  It seems that the  NICA JINR and FAIR GSI Projects  are  perfectly suited  to resolve  these tasks. 

\vspace*{2.2mm}

{\bf Acknowledgments.}
The authors  are   thankful to D. B. Blaschke and A. S. Sorin  for important  comments. 
The fruitful discussions with  S. V. Molodtsov,  D. H. Rischke and E. V. Shuryak
are  acknowledged.
This publication is based on the research provided by the grant support of the State Fund for Fundamental Research   (project N  {\cyrfnt F}58/175-2014).


\vspace*{0.22cm}



\begin{thebibliography}{99}

\bibitem{KABOliinychenko:12}
%
D.R. Oliinychenko, K.A. Bugaev and  A.S. Sorin, 
Ukr. J. Phys.  {\bf 58},  211 (2013). 


\bibitem{HRGM:13}
%
K. A. Bugaev, D. R. Oliinychenko, A. S. Sorin, G. M. Zinovjev,
Eur. Phys. J. A {\bf  49}, (2013) 30.

\bibitem{ResModif:12}
%
K. A. Bugaev, D. R. Oliinychenko,  E. G. Nikonov,   A. S. Sorin and G. M. Zinovjev, 
PoS Baldin ISHEPP XXI (2012) 017, 1-14;  arXiv:1212.0132  [hep-ph].
  
\bibitem{ResModif:13}
%
K. A. Bugaev, A. I. Ivanytskyi,  D. R. Oliinychenko, E. G. Nikonov, V. V. Sagun   and G. M. Zinovjev,
arXiv:1312.4367  [hep-ph].

\bibitem{KABAndronic:05}
%
A. Andronic, P. Braun-Munzinger and J. Stachel,  
Nucl. Phys. A {\bf 772}, 167 (2006).

\bibitem{KABAndronic:09}
%
A. Andronic, P. Braun-Munzinger and J. Stachel, Phys. Lett. B {\bf 673}, 142 (2009) and references therein.

\bibitem{Cleymans:06}
%
J. Cleymans, H. Oeschler, K. Redlich, and S. Wheaton, Phys. Rev. C 73, 034905 (2006).

\bibitem{KABCleymansFO}
%
%
J. Cleymans,  K. Redlich, Phys. Rev. Lett. {\bf 81}, 5284 (1998).

\bibitem{KABCleymansFO:b}
%
J. Cleymans, K. Redlich, Phys. Rev. C 61 054908 (1999). 

\bibitem{Tawfik:06a}
%
A. Tawfik, Europhys. Lett. {\bf 75}, 420 (2006).

\bibitem{Tawfik:06b}
%
A. Tawfik, Nucl. Phys. A {\bf 764}, 387 (2006).

\bibitem{EffEOS:73}
%
E. V. Shuryak, Sov. J. Nucl. Phys. {\bf 16},  220 (1973).

\bibitem{KABHagedorn4}
%
T. D. Cohen and V. Krejcirik,
  J.\ Phys.\ G {\bf 39}, 055001 (2012).

\bibitem{KABPDG:08}
%
C. Amsler {\it et al.}, Phys. Lett. B {\bf 667}, 1 (2008) [http://pdg.lbl.gov]. 

\bibitem{FWM:08}
%
K.A. Bugaev, V.K. Petrov and G.M. Zinovjev, 
Europhys. Lett. {\bf 85},  22002 (2009).


\bibitem{FWM:09}
%
K. A. Bugaev, V. K. Petrov and G. M. Zinovjev,
 Phys. Rev.  {\bf C  79},    (2009) 054913. 

\bibitem{KABGamow1}
%
C. S. Rolfs and W. S. Rodney, {\it Cauldrons in the Cosmos}, University of Chicago Press, 1986.

\bibitem{KABSigma:07}
%
R. Garcia-Martin, J. R. Pelaez and F. J. Yndurain,
Phys. Rev.  D {\bf 76},   074034  (2007).

\bibitem{KABStashek:87}
%
K. G. Denisenko and St. Mrowczynski, Phys. Rev. C {\bf  35},  1932 (1987). 

\bibitem{KABBarz:1991}
%
H.W. Barz {\it et al.},
Phys. Lett. B {\bf 265}, 219 (1991).

\bibitem{StrangSuppr:03}
%
F. Becattini {\it et al.}, Phys. Rev. C {\bf 69},   024905 (2004).

\bibitem{StrangSuppr:05}
%
J. Letessier, J. Rafelski, nucl-th/0504028. 

\bibitem{HRGM2:13}
%
K. A. Bugaev  {\it et al.}, 
 Europhys. Lett. {\bf 104} 22002   (2013).
  
 \bibitem{EarlyFO:1} 
%
K. A. Bugaev,
J. Phys. G {\bf 28},  1981 (2002).  



\bibitem{EarlyFO:2} 
%
 M. I. Gorenstein, K. A. Bugaev and  M. Gazdzicki,
Phys. Rev. Lett. {\bf 88}, 132301 (2002). 

\bibitem{EarlyFO:3} 
%
K. A. Bugaev, M. Gazdzicki, M. I. Gorenstein, 
Phys. Lett. B {\bf 544},  127 (2002).

\bibitem{Gupta13}
S. Chatterjee, R.M. Godbole, S. Gupta,
Phys. Lett. B, {\bf 727},  554 (2013).

\bibitem{SA1}
%
  K.~A.~Bugaev, A.~I.~Ivanytskyi, D.~R.~Oliinychenko, V.~V.~Sagun, I.~N.~Mishustin, D.~H.~Rischke, L.~M.~Satarov and G.~M.~Zinovjev,
  arXiv:1405.3575 [hep-ph].

 \bibitem{SA2}
  K.~A.~Bugaev, A.~I.~Ivanytskyi, D.~R.~Oliinychenko, V.~V.~Sagun, I.~N.~Mishustin, D.~H.~Rischke, L.~M.~Satarov and G.~M.~Zinovjev,
  arXiv:1412.0718 [nucl-th]. 


\end{thebibliography}
\end{document}